\documentclass{elsarticle}

\usepackage{graphicx}
\usepackage{amssymb,amsmath, amsthm}
\usepackage{multirow}

\begin{document}

\title{Unveiling saturation effects from nuclear structure function measurements at the EIC}

\author[Polytechnique]{Cyrille Marquet}

\author[USC]{Manoel R. Moldes\corref{correspondingauthor}}
\cortext[correspondingauthor]{Corresponding author}
\ead{manoel.rodriguez-moldes@usc.es}

\author[USC,BNL]{P\'\i a Zurita}

\address[Polytechnique]{Centre de Physique Th\'eorique, \'Ecole Polytechnique, CNRS, Universit\'e Paris-Saclay, 91128 Palaiseau, France}

\address[USC]{Departamento de F\'{i}õsica de Part\'{i}culas and IGFAE, Universidade de Santiago de Compostela, \\15706 Santiago de Compostela, Galicia-Spain}

\address[BNL]{Physics Department, Brookhaven National Laboratory, Upton, NY 11973, USA}


\begin{abstract}

We analyze the possibility of extracting a clear signal of non-linear parton saturation effects from future measurements of nuclear structure functions at the Electron-Ion Collider (EIC), in the small-$x$ region. Our approach consists in generating pseudodata for electron-gold collisions, using the running-coupling Balitsky-Kovchegov evolution equation, and in assessing the compatibility of these \emph{saturated} pseudodata with existing sets of nuclear parton distribution functions (nPDFs), extrapolated if necessary. The level of disagreement between the two is quantified by applying a Bayesian reweighting technique. This allows to infer the parton distributions needed in order to describe the pseudodata, which we find quite different from the actual distributions, especially for sea quarks and gluons. This tension suggests that, should saturation effects impact the future nuclear structure function data as predicted, a successful refitting of the nPDFs may not be achievable, which would unambiguously signal the presence of non-linear effects.

\end{abstract}



\maketitle

\section{Introduction}

It is foreseen that the next high-energy nuclear physics facility will be an Electron-Ion Collider (EIC) \cite{Accardi:2012qut}. One of the main physics goals of this future QCD laboratory will be to unambiguously unveil the onset of the so-called gluon saturation regime of QCD. This regime of hadronic and nuclear wave functions at small longitudinal momentum fraction $x$ is characterized by a transverse momentum scale, the saturation scale $Q_s(x)$, at which non-linearities become of comparable importance to linear evolution \cite{Gelis:2010nm}. The emergence of this non-linear regime is a fundamental consequence of QCD dynamics, and it has been subject to steady theoretical progress for the past twenty years. However the general consensus is that a clear experimental discovery remains elusive.

Saturation phenomenology has been successful in all the (admittedly small) phase-space corners where that physics is expected to be relevant (i.e., for every collider process that involves small-$x$ partons and transverse momenta of the order of the saturation scale), for a broad range of observables and various collision systems, including $d+\mathrm{Au}$ collisions at RHIC and in $e+p$ collisions at HERA \cite{Lappi:2010ek,Albacete:2014fwa}. The strongest hint in favor of saturation so far is arguably the suppression of the away-side peak of the di-hadron correlation function at forward rapidities in central $p+A$ collisions versus $p+p$ collisions \cite{Adare:2011sc,Braidot:2010zh}, discovered at RHIC after it had been predicted \cite{Marquet:2007vb}. Nevertheless, the global belief in the community is that a direct, unquestionable evidence is still missing, and that the EIC may provide it.

Given this past success, there is little doubt that saturation effects will be at play in $e+\mathrm{Au}$ collisions at the EIC. The question is which is the best way to uncover them, which are the golden observables and smoking-gun measurements. Measurements from which one can undoubtedly conclude that non-linear effects are clearly needed on top of standard linear perturbative QCD dynamics, but also from which one can thoroughly verify that those non-linear effects are under theoretical control, i.e., they are described by non-linear QCD equations as opposed to being fully of non-perturbative origin. In this work, we shall focus on the $F_2^{\mathrm{Au}}$ and $F_L^{\mathrm{Au}}$ structure functions at small $x$, whose combined measurement is often mentioned as the primary candidate to provide a smoking gun in favor of parton saturation \cite{Boer:2011fh}. They provide access to the distributions of quarks and gluons, respectively, and can be extracted from measuring the total $e+A \to e+X$ cross-section at different collision energies.

However, even if saturation effects would play an important role in the kinematical range of the future EIC data, there is a possibility that such effects would go unnoticed: standard QCD fits of parton distribution functions (PDFs), which are based on the linear DGLAP evolution equation and therefore do not take into account saturation effects, could still be able to describe structure function data inside the saturation regime \cite{Albacete:2012rx}. This is due to the large amount of freedom allowed in this approach for the initial conditions to the QCD evolution: those can be artificially suppressed in fits, in order to compensate for an evolution that would be too fast, as would be the case if the DGLAP equation was used into the saturation regime without the proper non-linear corrections.

In other words, up to a certain point, it is possible to obtain a successful fit of $F_2$ and $F_L$ data (and a resulting PDF set) with an incomplete evolution equation, which can unfortunately lead to wrong conclusions and an inaccurate description of the parton content. The question we set out to answer in this letter is whether or not saturation effects, as well as the lever arm in $x$, will be large enough in order to ensure that this will not happen with gold nuclei at the EIC, and that the combined measurement of the $F_2$ and $F_L$ structure functions is indeed a golden observable. We note that the possibility of detecting non-linear effects with inclusive measurements has been investigated in earlier works \cite{Eskola:2002yc,Cazaroto:2008iy,Accardi:2011qh}; the present work utilizes of a novel approach.

Our strategy is the following. We first generate $F_2^{\mathrm{Au}}$ and $F_L^{\mathrm{Au}}$ pseudodata, with saturation effects setting in according to the latest predictions based on the running-coupling Balitsky-Kovchegov (rcBK) evolution \cite{Balitsky:1995ub,Kovchegov:1999yj,Gardi:2006rp,Kovchegov:2006vj,Balitsky:2006wa,Balitsky:2008zza}. Then we answer the question whether various nuclear PDF (nPDF) fits on the market (namely\footnote{The choice of these two sets obey to practical reasons. Though nowadays there are several sets of nPDFs available, EPS09 presents the most extreme effect on the gluon density, while DSSZ shows almost no difference between the proton and nuclear gluon. Thus, by considering these two sets, we cover all possible spectrum of the known nPDFs.} EPS09 \cite{Eskola:2009uj} and DSSZ \cite{deFlorian:2011fp}), which are well constrained by large-$x$ large-$Q^2$ existing data, can accommodate our EIC pseudodata. This is done using a Bayesian reweighing procedure which addresses the compatibility of new data (and the quantitative modifications they might induce) to an existing set of PDFs \cite{Giele:1998gw,Ball:2010gb,Ball:2011gg,Watt:2012tq,Watt:2013oha,Sato:2013ika,Paukkunen:2014zia}. If the answer is positive, then saturation effects at the EIC would stay concealed; if the answer is negative, then non-linear QCD dynamics would be needed to describe the data. In the former case, measuring deep inelastic scattering (DIS) off heavy nuclei would still be crucial in order to extract nuclear modifications of quark and gluon distributions at generic values of $x$, but the structure functions would not form smoking-gun measurements for parton saturation.


The plan of the letter is as follows. In section \ref{sec:saturation}, we detail the saturation framework used to generate our structure-function pseudodata for gold nuclei at the EIC. In section \ref{sec:Bayes}, we explain the most relevant features of the the Bayesian reweighing procedure used to check the compatibility of our pseudodata with the nPDFs sets. The obtained results are the topic of Sec. \ref{sec:Results}. Finally we summarize our findings in Sec. \ref{sec:summary}.

\section{Nuclear structure function pseudodata from the rcBK saturation model}\label{sec:saturation}

It is customary to write the DIS structure functions $F_2$ and $F_L$ as follows:
\begin{equation}
\begin{split}
F_2(x,Q^2)& =\frac{4\pi^2\alpha_{em}}{Q^2}\left[\sigma^{\gamma^*A}_{T}(x,Q^2)+\sigma^{\gamma^*A}_{L}(x,Q^2)\right] \, , \\
F_L(x,Q^2)&=\frac{4\pi^2\alpha_{em}}{Q^2}\sigma^{\gamma^*A}_{L}(x,Q^2)\ ,
\end{split}
\end{equation}
where $\sigma^{\gamma^*A}_{L,T}$ represents the $\gamma^*$+A total cross-section, for a transversely (T) or longitudinally (L) polarized virtual photon.
In turn, at small $x$, these are obtained in the following way:
\begin{equation}\label{eq:sigmatot}
\sigma^{\gamma^*A}_{L,T}(x,Q^2)
= \int d^2r \int_0^1 dz \left| \Psi^{\gamma^*}_{L,T}(z,{\bf r};Q^2)\right|^2 \int d^2b \frac{d\sigma^A_\textrm{dip}}{d^2b}({\bf r},{\bf b};x)\ ,
\end{equation}
where $\left| \Psi_{L,T}^{\gamma^*}(z,{\bf r};Q^2) \right|^2$ represents the probability for a photon of virtuality $Q^2$ to produce a $q\bar q$ pair 
of transverse size ${\bf r}$ and $z$ denotes the fraction of longitudinal momentum carried by the quark.
$d\sigma^A_\textrm{dip}/d^2b$ stands for the total cross-section for this $q\bar q$ pair to scatter off the target nucleus at an impact parameter ${\bf b}$ and contains the QCD dynamics. By contrast, $\Psi_{L,T}^{\gamma^*}$ is well known from QED.

To compute the $q\bar q$ dipole-nucleus cross-section, we follow the approach of \cite{Armesto:2002ny,Kowalski:2003hm}. Using the optical theorem, the total cross-section can be written $d\sigma^A_\textrm{dip}/d^2b=2(1-S_A)$ where the $S$-matrix element $S_A$ represents the probability amplitude for the dipole to \emph{not} interact with the target nucleus. Then, the essence of the model is to write that the dipole scatters independently off the different nucleons. Introducing the coordinates of the individual nucleons $\left\{{\bf b}_i\right\}$, this means $S_A({\bf r},{\bf b};x) = \prod_{i=1}^A S_p({\bf r},{\bf b}-{\bf b}_i;x)$, where $S_p$ refers to the $q\bar q$ dipole-nucleon $S$-matrix element. 

We further assume that the positions of the nucleons $\left\{{\bf b}_i\right\}$ are distributed according to the Woods-Saxon distribution $T_A({\bf b})$:
\begin{equation} \label{eq:defTA}
T_A({\bf b}) = \int dz \frac{C}{1+\exp \left[\left( \sqrt{{\bf b}^2 + z^2} - R_A \right)/d \right]}\ ,
\end{equation}
which is normalized to unity $\int d^2b\ T_A({\bf b}) = 1$. The nuclear radius
$R_A$ and surface diffuseness $d$ are measured  from the electric charge distribution, their values can be found in Ref.~\cite{DeJager:1987qc}.

In this model, the $q\bar q$ dipole-nucleus cross-section can therefore be written
\begin{equation}
\frac{d\sigma^A_\textrm{dip}}{d^2b}({\bf r},{\bf b};x) = \int \prod_{i=1}^A \left\{ d^2b_i T_A({\bf b}_i)\right\} 2\left[1-\prod_{i=1}^A S_p({\bf r},{\bf b}-{\bf b}_i;x)\right]\ ,
\label{eq:nuc-average}
\end{equation}
which is well approximated by \cite{Kowalski:2003hm} 
\begin{equation}
\frac{d\sigma^A_\textrm{dip}}{d^2b} \approx 2\left[1-\left(1-\frac{T_A({\bf b})}{2}\sigma^p_\textrm{dip} \right)^A\right]\ .
\label{eq:nuc-dipole}
\end{equation}
The expression in parenthesis in \eqref{eq:nuc-dipole} can be further replaced by $\exp\left(-AT_A({\bf b})\sigma^p_\textrm{dip}/2\right)$ for large $A$.
The parameters of the model come from the Woods-Saxon distribution \eqref{eq:defTA} and from the AAMQS fit \cite{Albacete:2010sy} of the $q\bar q$ dipole-nucleon
cross-section $\sigma^p_\textrm{dip}$ to the reduced cross-section data from HERA, which we shall detail now.

In the AAMQS fit, the dipole cross-section is expressed in terms of an impact-parameter independent dipole scattering amplitude ${\cal N}$, which contains the small-$x$ QCD dynamics:
\begin{equation}
\sigma^p_\textrm{dip}({\bf r},x)=\sigma_0 {\cal N}({\bf r},x)\ ,
\end{equation}
We note that other models could be used here, in particular models that implement a non-trivial impact parameter dependence already at the level of dipole-nucleon
cross section, however, this does not lead to important differences for the inclusive observables considered in this work. It would in the case of diffractive and exclusive observables, but this goes beyond the scope of this paper.

The dipole scattering amplitude ${\cal N}$ obeys the rcBK equation \cite{Balitsky:1995ub,Kovchegov:1999yj,Gardi:2006rp,Kovchegov:2006vj,Balitsky:2006wa,Balitsky:2008zza}:
\begin{equation}
  \frac{\partial {\cal N}(r,Y)}{\partial Y}=\int d^2{\bf r_1}\,
  K^{{\rm run}}({\bf r},{\bf r_1},{\bf r_2}) \left[{\cal N}(r_1,Y)+{\cal N}(r_2,Y)-{\cal N}(r,Y)- {\cal N}(r_1,Y)\,{\cal N}(r_2,Y)\right]\ ,
\label{eq:bk1}
\end{equation}
with $Y=\ln(x_0/x)$. Using Balitsky's prescription \cite{Balitsky:2006wa}, the kernel in \eqref{eq:bk1} reads
\begin{equation}
  K^{{\rm run}}({\bf r},{\bf r_1},{\bf r_2})=\frac{N_c\,\alpha_s(r^2)}{2\pi^2}
  \left[\frac{r^2}{r_1^2\,r_2^2}+
    \frac{1}{r_1^2}\left(\frac{\alpha_s(r_1^2)}{\alpha_s(r_2^2)}-1\right)+
    \frac{1}{r_2^2}\left(\frac{\alpha_s(r_2^2)}{\alpha_s(r_1^2)}-1\right)
  \right]\,,
\label{kbal}
\end{equation}
where ${\bf r_2}={\bf r}-{\bf r_1}$ and $v\equiv |{\bf v}|$ for two-dimensional vectors. Following \cite{Albacete:2009fh}, the running coupling is regulated in equations \eqref{eq:bk1} and \eqref{kbal} by freezing it to a constant value $\alpha_s^{fr}=0.7$ in the infrared. The initial condition for the evolution is the so-called McLerran-Venugopalan (MV) model \cite{McLerran:1997fk}:
\begin{equation}
\mathcal{N}_{F}(r,Y=0)=1-\exp\left[ -\frac{r^2\,Q_{s0}^2}{4}\,\ln\left(\frac{1}{\Lambda\,r}+e\right)\right]\ ,
\end{equation} 
where $Q_{s0}^2$ is the initial saturation scale, and where $\Lambda=0.241$ GeV.
All the parameters can be found in \cite{Albacete:2010sy}, we shall make use of the ``e" fit of that reference. Recently, more advanced fits have appeared where Eq.~\eqref{eq:bk1} is also supplemented with collinear resummations \cite{Albacete:2015xza}; however those are not yet available for public use, but could as easily be used in the future.

\begin{figure}[t]
\includegraphics[width=\textwidth]{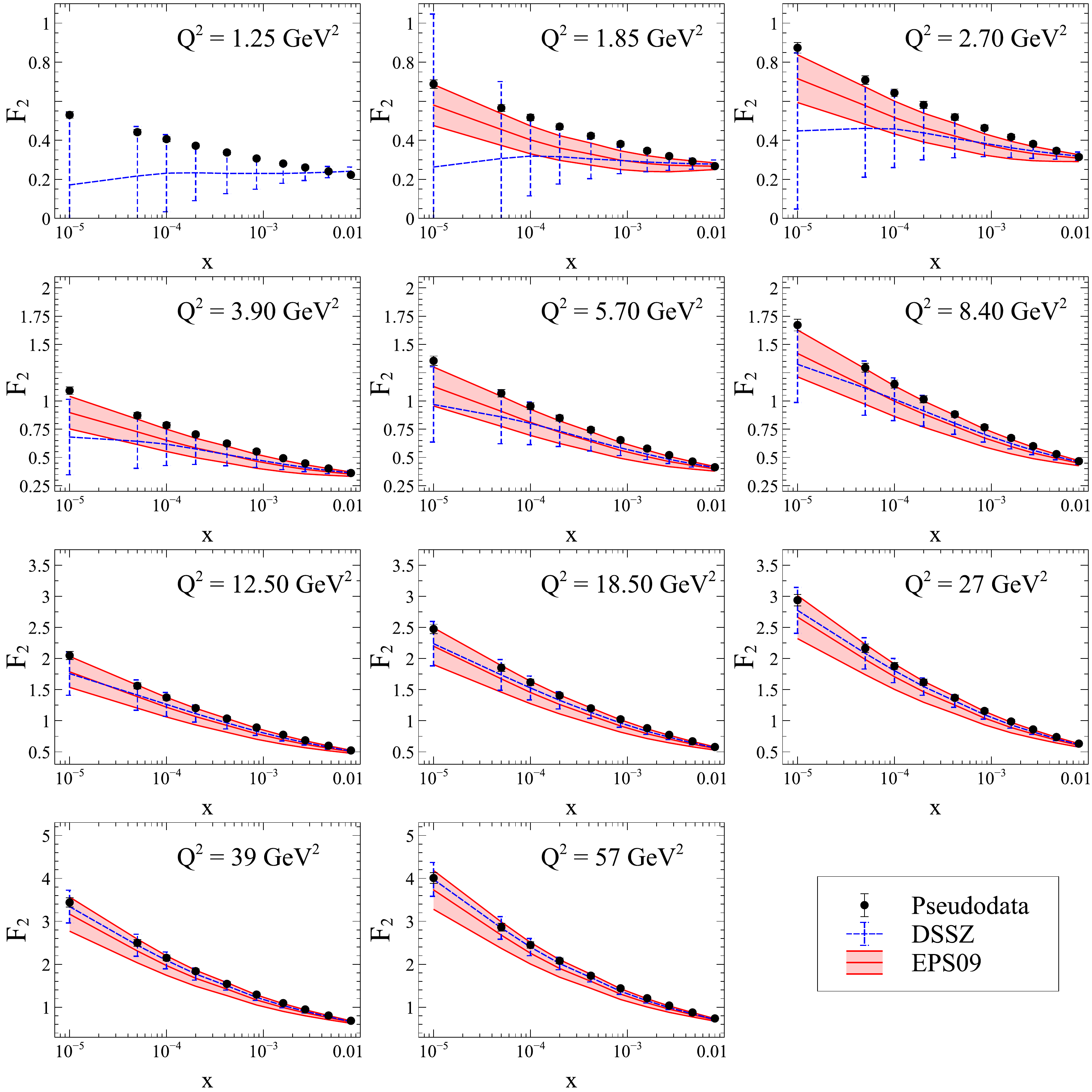}
\caption{$F_{2}^\textrm{Au}$ pseudodata (normalized to the number of nucleons $A = 197$) obtained from the rcBK saturation model, and the corresponding values computed in the collinear factorization framework with the EPS09 and DSSZ nPDFs.}
\label{fig:F2_data}
\end{figure}

\begin{figure}[htbp]
\begin{center}
\includegraphics[width=\textwidth]{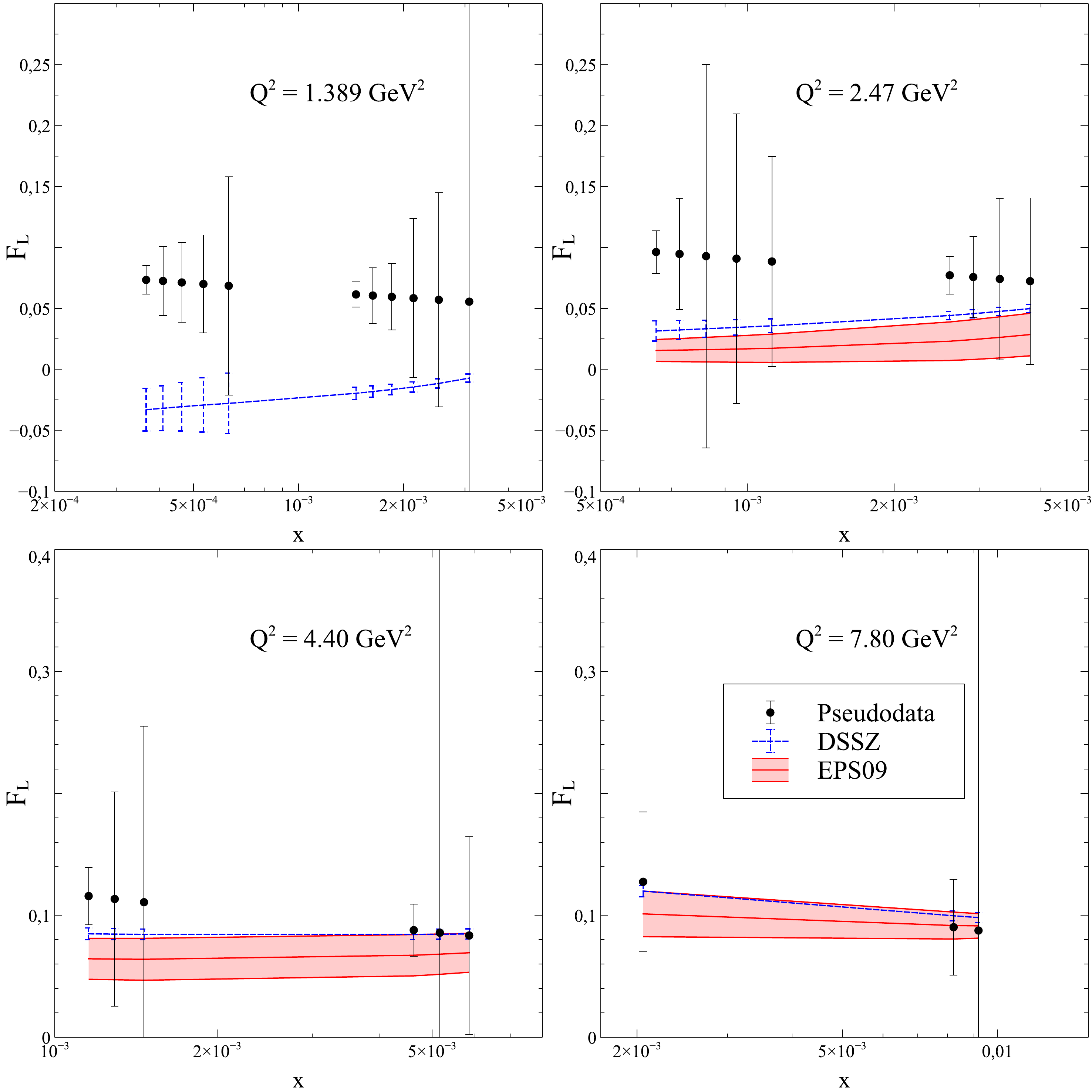}
\caption{$F_{L}^\textrm{Au}$ pseudodata (normalized to the number of nucleons $A = 197$) obtained from the rcBK saturation model, and the corresponding values computed in the collinear factorization framework with the EPS09 and DSSZ nPDFs.}
\label{fig:FL_data}
\end{center}
\end{figure}

With all the ingredients described above, we can generate the central values of our pseudodata for $F_2^\textrm{Au}$ and $F_L^\textrm{Au}$. Concerning the projected error bars, they have been generated for us by the ``BNL Task Force for the EIC" \cite{Matt}. As in many EIC dedicated studies, an integrated luminosity of $10 fb^{-1}$ was assumed, implying that systematic uncertainties dominate. The latter are estimated to be around $3\%$ for $F_2$, and for $F_L$, three centre-of-mass energies ranging from 30 to 90 GeV were assumed for the extraction, yielding bigger errors. In Figs. \ref{fig:F2_data} and \ref{fig:FL_data}, we show a comparison of those pseudodata, with calculations obtained from the EPS09 and DSSZ nPDF sets, with $x$ ranging from $10^{-5}$ to $10^{-2}$, with $Q^{2}$-values within the perturbative QCD range. These theoretical curves are the result of incorporating the nuclear PDFs into the next-to-leading order code for structure functions from Ref.~\cite{Martin:2009iq}, meaning that we are using the same MSTW proton baseline for both our EPS09 and DSSZ nPDF calculations.


In the case of $F_{2}$ (Fig.~\ref{fig:F2_data}), the pseudodata at the highest $x$-values are well within the theoretical uncertainties for both nPDF sets considered. In contrast, for $x \lesssim 10^{-3}$ the rcBK prediction lies in the upper limit of the nPDFs uncertainties or even above them. This is due to the fact that nPDFs fits are, at the moment, not constrained at such low $x$-values, and the theoretical predictions shown on the figures rely on extrapolations. Moreover, contrary to what one could have naively expected, the nPDF extrapolations induce $F_2$ values which are smaller than what the saturation model predicts, not larger. This means that, with the current nPDF sets, one should not expect the {\it golden} scenario in which the collinear factorization predictions will overshoot the data at small $x$, due to the lack of non-linear effects in that framework.

With respect to the longitudinal structure function, there are two main reasons that explain the extreme differences seen in Fig.~\ref{fig:FL_data}. On the one hand, as before, part of the generated pseudodata lie in an unexplored kinematical region and thus the nPDF extrapolations are not reliable. On the other hand, $F_{L}$ is sensitive to the gluon distribution already at the lowest order, unlike $F_{2}$. This gluon density is not well determined, as the bulk of data considered in nPDFs extractions is not sensitive to it, either due to the kinematical range covered or to the observables themselves.


\begin{figure}[htbp]
\begin{center}
\includegraphics[width=\textwidth]{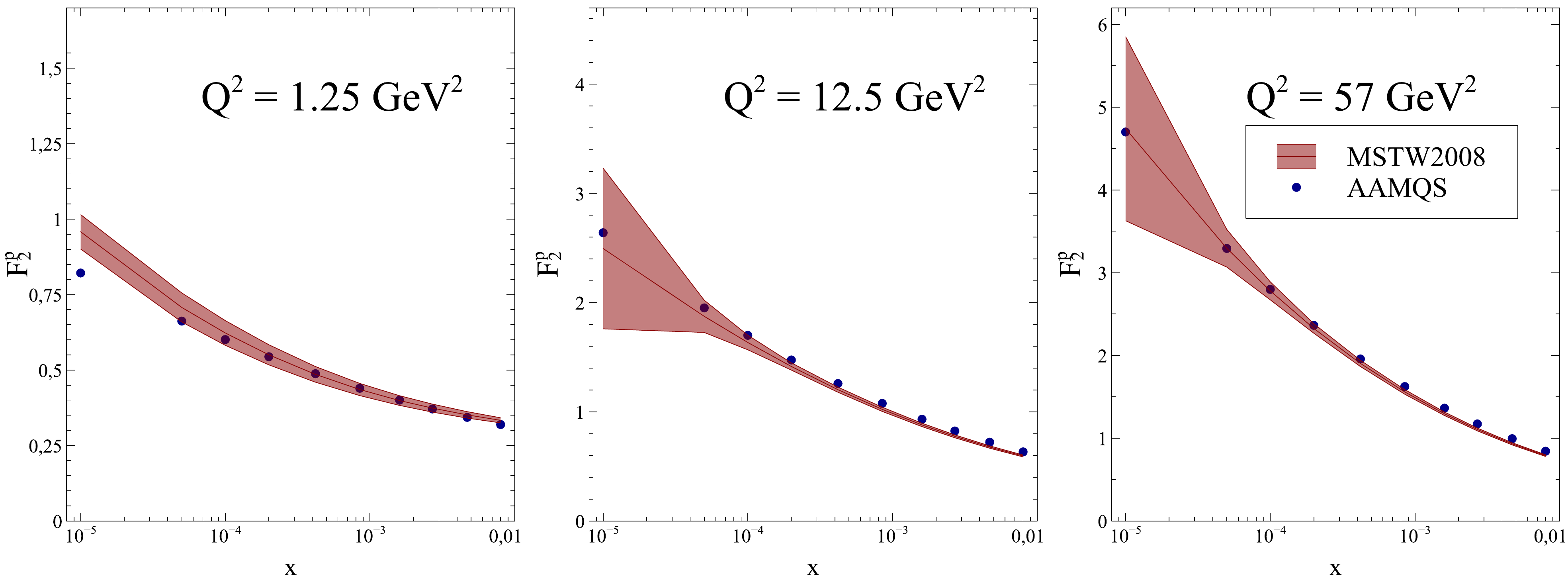}
\caption{Comparison of $F_{2}^\textrm{p}$ predictions obtained from the rcBK saturation model and the collinear-factorization framework with the MSTW PDFs.}
\label{fig:F2_proton}
\end{center}
\end{figure}

All these reasons come together to explain the level of disagreement between the \emph{saturated} pseudodata and the collinear factorization predictions. And to illustrate that these discrepancies relate to nuclear effects, and not to differences at the level of the proton structure functions, we show in Fig.~\ref{fig:F2_proton} that indeed the AAMQS and the collinear-factorization calculations are compatible, as they should since these are build upon the same HERA data. 

Finally, it is noteworthy that the upper left panel in both Figs.~\ref{fig:F2_data} and \ref{fig:FL_data}, corresponding to the smallest $Q^{2}$-value, only has a prediction for DSSZ, as it is outside the validity range of EPS09. Given this, the total number of pseudodata included in the analysis below is not the same for each nPDF set. We would also like to point out that the projected error bars for the $F_L$ measurements at the EIC are much bigger than those in the case of $F_2$, and as a consequence including our $F_L$ pseudodata in the analysis makes little difference.

\section{Bayesian reweighting of nuclear PDFs}\label{sec:Bayes}

The partonic densities (PDFs) are a necessary piece for the theoretical predictions of physical observables for processes involving at least one hadron in the initial or final state. However, their determination from experimental data is an involved and time consuming procedure. For this reason, updating the PDFs every time a new set of data becomes available is rather vexing, as a priori one can't know if new information would be obtained from it.

Therefore, statistical methods have been developed in order to bypass this obstacle: the reweighting techniques are tools that allow to incorporate information from newly measured data into a set of PDFs without recurring to the standard procedure of global fitting. At least two methods are now available: the Hessian reweighting and the Bayesian reweighting \cite{Giele:1998gw,Ball:2010gb,Ball:2011gg,Watt:2012tq,Watt:2013oha,Sato:2013ika,Paukkunen:2014zia}. While in this work we chose the latter, it has been shown that they are equivalent for PDF sets with theoretical uncertainties determined by the Hessian method as the ones we use.

Let us assume that we have the representation of the underlying probability distribution $\mathcal{P}_{\rm old}(f)$ of the PDFs given by a large ensemble of PDFs $f_k$, $k=1 \ldots N_{\rm rep}$. Then the expectation value for any quantity $\mathcal{O}$ depending on the PDFs can be computed as 
\begin{equation}\label{eq:obs}
\langle \mathcal{O} \rangle = \frac{1}{N_{\rm rep}} \sum_{k=1}^{N_{\rm rep}} \mathcal{O} \left[ f_k \right] \, , 
\end{equation}
with variance 
\begin{equation}\label{eq:varold}
\delta \langle \mathcal{O} \rangle = \sqrt{\frac{1}{N_{\rm rep}} \sum_{k=1}^{N_{\rm rep}} \left( \mathcal{O} \left[ f_k \right] - \langle \mathcal{O} \rangle\right)^2 } \, .
\end{equation}

If new data $\vec{y}\equiv \lbrace y_{i}, i=1, ..., N_{\mathrm{data}} \rbrace$ with covariance matrix $C_{ij}$ are made available, we can update $\mathcal{P}_{\rm old}(f)$ incorporating the information contained in the new data by use of the Bayes theorem as 
\begin{equation}
\mathcal{P}_{\rm new}(f) \varpropto \mathcal{P}(\vec y \vert f) \, \mathcal{P}_{\rm old}(f)\, ,
\label{eq:proba}
\end{equation}
where $\mathcal{P}(\vec y  \vert f)$ is the likelihood for the new data given the original set of parton densities. With this modification the quantities defined in equations \eqref{eq:obs} and \eqref{eq:varold} turn into weighted averages 
\begin{eqnarray}
\langle \mathcal{O} \rangle_{\rm new}  & = & \frac{1}{N_{\rm rep}} \sum_{k=1}^{N_{\rm rep}} \omega_k \, \mathcal{O} \left[ f_k \right]\ , \label{eq:Bnew} \\
\delta \langle \mathcal{O} \rangle_{\rm new}  & = & \sqrt{\frac{1}{N_{\rm rep}} \sum_{k=1}^{N_{\rm rep}} \omega_k \, \left( \mathcal{O} \left[ f_k \right] - \langle \mathcal{O} \rangle_{\rm new} \right)^2 } \ , \label{eq:Bnew2}
\end{eqnarray}
with the weights $\omega_{k}$ proportional to the likelihood.

The adequate selection of the likelihood is a delicate matter and several options have been proposed \cite{Sato:2013ika,Giele:1998gw,Ball:2010gb,Ball:2011gg}. In particular, we follow the option of \cite{Paukkunen:2014zia}, equivalent to a refit for PDF sets with uncertainties based on the Hessian method (with $N_{\rm eig}$ eigenvalues and fixed tolerance $\Delta \chi^2$):
\begin{equation}
\omega_k
= \frac{\exp\left[-\chi^2_k/2\Delta \chi^2\right]}{(1/N_{\rm rep}) \sum_{k=1}^{N_{\rm rep}} 
\exp\left[-\chi^2_k/2\Delta \chi^2\right]}\ , \label{eq:pesos}
\end{equation}
where the theoretical values $y_i \left[f \right]$ are estimated by
\begin{equation}
y_i \left[f \right] \approx y_i \left[{S_0} \right] + \sum_k \frac{\partial y_i \left[{S} \right]}{\partial z_k}{\Big|_{S=S_0}} z_k \ ,
\end{equation}
with the deviation between data and each replica $k$ computed as
\begin{equation}
 \chi^{2}_{k} = \sum_{i,j=1}^{N_{\rm data}} \left(y_i[f_k]-y_i\right) C_{ij}^{-1} \left(y_j[f_k]-y_j\right) \, . \label{eq:chi2onlynew}
\end{equation}
An interesting feature of the Bayesian method is the existence of a quantitative estimator of the agreement between the data originally considered for the PDF fit and the new one. This estimator, the effective number of replicas $N_{\rm eff}$, is defined as
\begin{equation}
N_{\rm eff} \equiv \exp \left\{ \frac{1}{N_{\rm rep}} \sum_{k=1}^{N_{\rm rep}} \omega_k \log(N_{\rm rep}/\omega_k)\right\} \, .
\end{equation}
$N_{\rm eff} \ll N_{\rm rep}$ points to a large tension between data sets, either due to incompatibility or too much new information in the new data. $N_{\rm eff} \approx N_{\rm rep}$ instead hints at a strong compatibility and the use of the new reweighted PDF set reliable. Nevertheless, the real meaning of $N_{\mathrm{eff}}$ can be less clear when comparing different PDFs sets, as was pointed in \cite{Armesto:2015lrg}.

The most general version of the procedure, applicable to any process with at least one hadron (or nucleus) in the initial or final state, starts by generating the replicas $f_k$ by
\begin{equation}
 f_{k} \equiv f_{S_0} + \sum_i^{N_{\rm eig}} \left( \frac{f_{S^+_i}-f_{S^-_i}}{2} \right) R_{ik} \label{eq:replicas} \, ,
\end{equation}
with $ f_{S_0}$ and ${f_{S^{\pm}_i}}$ the PDFs for the central fit and eigenvectors, respectively, and $R_{ik}$ random numbers with Gaussian distribution centered at zero and with variance one. Once these are calculated for a high enough amount of replicas ($N_{\mathrm{rep}} > 10^{3}$), they are used to compute the theoretical values of the observable needed for each replica. For most of the  studied quantities it involves a large quantity of time consuming convolutional integrals and this goes in detriment of the reweighting positive feature of speed. However, should the PDF set under study enter the computation linearly (as in the present case), it is possible to alter the order of the steps and save time. Let us consider the electron-nucleus collision we are investigating. For the $k-th$ replica, the observable can be schematically written as 
\begin{equation}
\mathcal{O}_{k}= \hat{\mathcal{O}} \otimes f^{\rm A}_{k} \, ,
\end{equation}
where $\otimes$ denotes the convolution of the hard part $\hat{\mathcal{O}}$ with the PDF, and the sum over the partonic species is implicit. Using Eq. (\ref{eq:replicas}) to replace $f^{\rm A}_{k}$, we end up with
\begin{equation}
\mathcal{O}_{k}= \hat{\mathcal{O}} \otimes \left[f^{\rm A}_{S_0} + \sum_{i}^{N_{\rm eig}} \left( \frac{f^{\rm A}_{S^{+}_i}-f^{\rm A}_{S^{-}_i}}{2} \right) R_{ik}\right]  \, .
\end{equation}
By distributing the terms we have
\begin{equation}
\mathcal{O}_{k} = \mathcal{O}_{S_0} + \sum_{i}^{N_{\rm eig}} \frac{R_{ik}}{2} \left[ \mathcal{O}_{S_i^+} - \mathcal{O}_{S_i^-} \right],
\end{equation}
where $\mathcal{O}_{S_0}$ is the observable obtained with the central set, and $\mathcal{O}_{S_i^\pm}$ the ones corresponding to the eigenvectors. Thus we avoid computing the observable $N_{\rm rep}$ times and only do so $2N_{\rm eig}+1$ times.

\section{Results}
\label{sec:Results}


Given the pseudodata of Figs. \ref{fig:F2_data} and \ref{fig:FL_data} for $e+\mathrm{Au}$ collisions ($A = 197$), we perform a Bayesian reweighting using an initial number of  replicas $N_{\mathrm{rep}}=10^5$. For the sake of making better sense of the results, we perform the analysis three times, considering only $F_2$, only $F_L$ and both $F_2$ and $F_L$ pseudodata together. The results are shown on Table \ref{table:Results}, which summarizes the quantitative estimators of the adequacy of the nPDF description.



\begin{table}[t]
\centering
\begin{tabular}{c|c|ccc}
Pseudodata & nPDF & $\chi^2/n|_{\mathrm{before}}$ & $\chi^2/n|_{\mathrm{after}}$ & $N_{\mathrm{eff}}$ \\ \hline
\multirow{2}{*}{$F_2$} & DSSZ & 84.22 & 2.24 & 1877 \\
 & EPS09 & 26.51 & 1.38 & 15010 \\ \hline
\multirow{2}{*}{$F_L$} & DSSZ & 197.63 & 162.96 & 74335 \\
 & EPS09 & 42.03 & 39.05 & 98045 \\ \hline
\multirow{2}{*}{$F_2$ + $F_L$} & DSSZ & 109.06 & 38.15 & 1865 \\
 & EPS09 & 29.66 & 5.67 & 14910
\end{tabular}%
\caption{Results of the reweighting process: pseudodata taken into account (first column), nPDF considered (second column), $\chi^2$ per number of pseudodata points before and after the reweighting (third and forth column, respectively), and effective number of replicas remaining (fifth column) of the $N_{\mathrm{rep}}=10^5$ initial replicas.}
\label{table:Results}
\end{table}

If we look at the central column of Table \ref{table:Results}, we see that the $\chi^2$ is always larger for DSSZ than for EPS09. This is because, as mentioned before, the pseudodata in the upper left panel of Fig. \ref{fig:F2_data} and \ref{fig:FL_data} is only included in the analysis with the DSSZ set. This fact is specially relevant in the case of $F_L$, since the discrepancy between pseudodata and theoretical curve is larger than for $F_2$.  \\
Regarding $F_{2}$, the deviation of the pseudodata from the central predictions combined with the small uncertainties give huge contributions to the $\chi^{2}$ (central column).
As for $F_{L}$, we have to distinguish between nPDFs, in spite of the description not being good in either case. The error bars are much larger than for $F_2$, so this yields a total $\chi^2$ per number of points lower for EPS09. Nevertheless, for DSSZ, the inclusion of the upper left panel of Figure \ref{fig:FL_data} means a towering increase of $\chi^2$, despite the larger error bars.

These $\chi^{2}$ determine the weights according to equation \eqref{eq:pesos} which in turn give us the total number of meaningful replicas ($N_{\mathrm{eff}}$) that we see in the far right column of Table \ref{table:Results}. If we compare $N_{\mathrm{eff}}$ with the total number of replicas $N_{\mathrm{rep}}$, we can see, on one hand, that for $F_L$ most of the replicas survive (for EPS09, the practical totality of the replicas survive), noting that this pseudodata is compatible with the current nPDF fits. This does not come as a suprise, given the huge error bars of the $F_L$ pseudodata. On the other hand, considering both $F_2$ and $F_L$, we can see that less than $15\%$ of the replicas survive the reweighting in the case of EPS09. This number shrinks to less than $2\%$ if we consider the DSSZ set.  This means that there is a big tension between the pseudodata and the theoretical predictions, and a new fit should be mandatory.

\begin{figure}[bp]
\includegraphics[width=\textwidth]{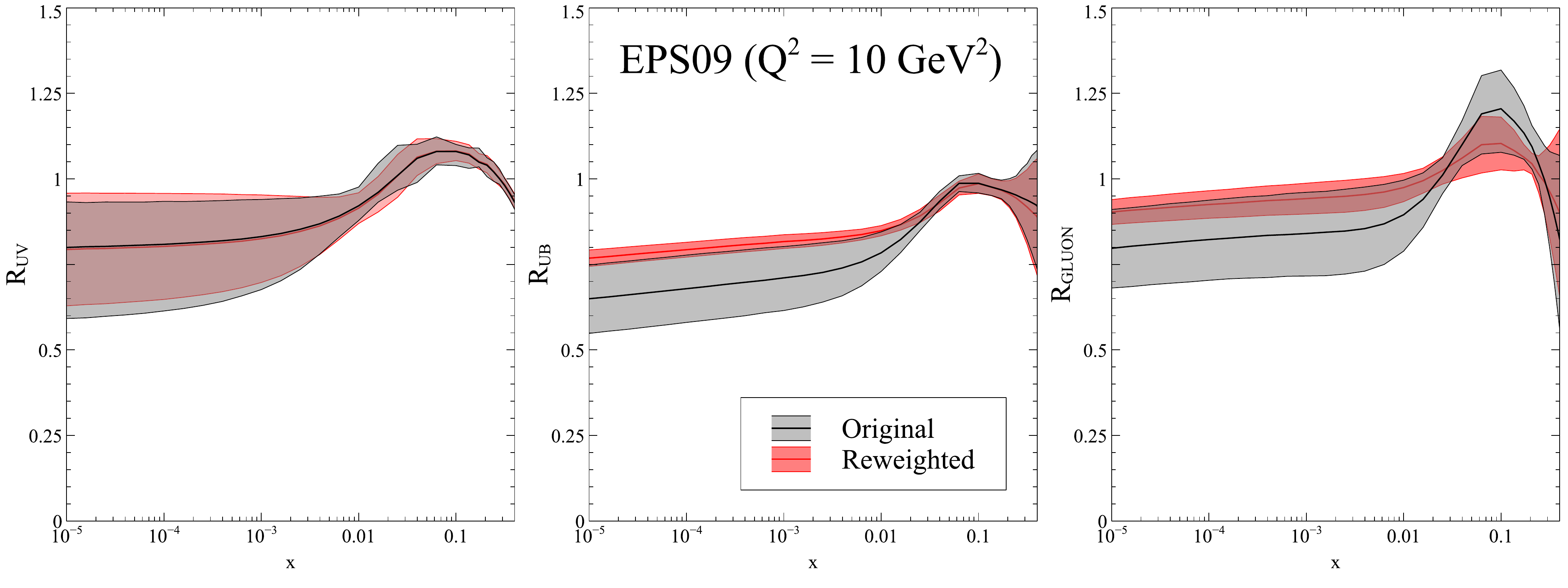}
\caption{$u$ valence quark (left), $\bar{u}$ sea quark (center) and gluon (right) distributions for the EPS09 PDF set before and after the reweighting.}
\label{fig:eps09_rew}
\end{figure}

Nevertheless, we can study what happens with the few remaining replicas and what nuclear partonic behavior is favored by the pseudodata. The reweighting process affects the nPDFs in the following way:

--- In the case of EPS09 (Fig.~\ref{fig:eps09_rew}), the reweighting suppresses minimally the central value of the valence at low $x$ (left panel), slightly increasing the shadowing in that region while the uncertainty is shifted up just a little bit. For the sea distribution (central panel) the central value increases and the shadowing suppression gets smaller, with the uncertainties shrinking dramatically. Finally the gluon density (right panel) flattens and the strong shadowing/anti-shadowing that characterizes this fit smoothes so much that leaves a curve almost compatible with unity.


\begin{figure}[htbp]
\includegraphics[width=\textwidth]{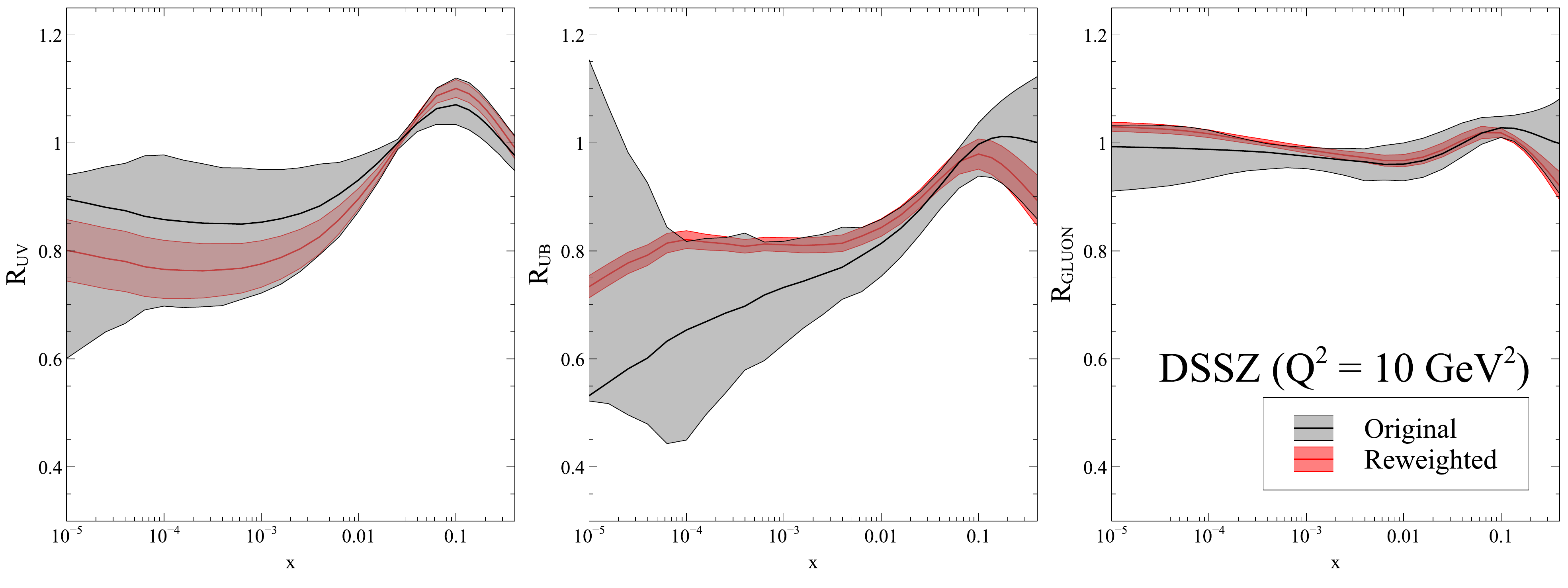}
\caption{$u$ valence quark (left), $\bar{u}$ sea quark (center) and gluon (right) distributions for the DSSZ PDF set before and after the reweighting.}
\label{fig:dssz_rew}
\end{figure}

--- In the case of DSSZ (Fig.~\ref{fig:dssz_rew}), the reweighting returns a valence distribution with stronger shadowing and anti-shadowing regions, a sea distribution with an enhancement of the central value in the shadowing area and a deeper EMC-effect. The reweighted gluon distribution behaves the same way as the reweighted sea, but the shadowing region for the gluon goes over unity, becoming, in fact, anti-shadowing. All these distributions get narrower uncertainties after the reweighting.

The curious behavior at low $x$ produced by the pseudodata originates from the ansatz in the initial parameterizations. In EPS09, the limit of $R$ for $x \to 0$ is a parameter, while for DSSZ there is no such constrain. Once obtained a fit, any prediction outside the kinematical range probed comes from an extrapolation, thus allowing for the puzzling and unrealistic curves at small $x$ of Fig. \ref{fig:dssz_rew} to occur.   

\section{Summary}
\label{sec:summary}

On the search for gluon saturation we have analysed the impact of future nuclear structure function low-$x$ data at the EIC. Our aim was to assess whether or not the collinear factorization approach with nuclear PDFs will be able to fit the data, if saturation sets in according to current expectations. As it is customary for nuclear PDFs, in this study we used the structure function $F_{2}$. In addition, we have also considered the longitudinal structure function $F_L$, as it is much more sensitive to the gluon density and thus potentially capable of providing interesting information, although with hindsight we found it not to be the case due to the large projected errors.

We generated  {\it saturated} pseudodata for $F_2^\textrm{Au}$ and $F_L^\textrm{Au}$, with central values obtained from the rcBK predictions and error bars estimated taking into account the design parameters the future detectors, and compared them with the corresponding collinear factorization predictions. The latter, that rely on extrapolations as most of the points lie outside the kinematical range explored in the original fits, lead to smaller structure functions than what the saturation model predicts at small $x$. This is particularly true for $F_L$, as the extrapolation for the nuclear gluon distributions at low $x$ are very unreliable. The quantitative impact of the pseudodata on the nPDFs was obtained by means of a Bayesian reweighting technique. The results look quite different from the original distributions, especially for sea quarks and gluons. This strong tension is confirmed by the numbers in Table \ref{table:Results}. The effect is bigger in the case of DSSZ for which the gluon parameterization lacks flexibility, but numerical estimators confirm the existence of a non trifling tension also in the case of EPS09. 

Our results suggest that, should the EIC provide data compatible with the expected theoretical description from the saturation model studied (or a similar one), a successful refitting of the nPDFs may not be achievable, which would unambiguously signal the presence of non-linear effects. However, in order to be fully conclusive and determine whether or not genuine saturation effects can be unveiled from nuclear structure function measurements, performing a new global nPDF fit will be necessary. At the moment we can not exclude the possibility of a successful refitting of the nuclear PDFs because the nuclear gluon distribution is currently essentially unconstrained at small-$x$. In that case, one would have to resort to diffractive observables in order to pin down saturation effects at the EIC (see e.g. \cite{Kugeratski:2005ck,Kowalski:2007rw,Kowalski:2008sa}).

\section*{Acknowledgements}

M.R.M. thanks the hospitality of the CPTh at \'Ecole Polytechnique, where part of this work was performed.
This research was supported by the European Research Council grant HotLHC ERC-2011-StG-279579; Ministerio de Ciencia e Innovaci\'on of Spain under project FPA2014-58293-C2-1-P; Xunta de Galicia (Conseller\'{\i}a de Educaci\'on) --- the group is part of the Strategic Unit AGRUP2015/11. P.Z. was supported in part by the Office of Nuclear Physics within the U.S. DOE Office of Science. M.R.M. was supported by Spanish INEM (SEPE).


\section* {References}

\begin{thebibliography}{99}

\bibitem{Accardi:2012qut}
  A.~Accardi {\it et al.},
  Eur.\ Phys.\ J.\ A {\bf 52} (2016) no.9,  268.
  [arXiv:1212.1701 [nucl-ex]].
  
\bibitem{Gelis:2010nm}
  F.~Gelis, E.~Iancu, J.~Jalilian-Marian and R.~Venugopalan,
  Ann.\ Rev.\ Nucl.\ Part.\ Sci.\  {\bf 60} (2010) 463.
  [arXiv:1002.0333 [hep-ph]].

\bibitem{Lappi:2010ek}
  T.~Lappi,
  Int.\ J.\ Mod.\ Phys.\ E {\bf 20} (2011) 1.
  [arXiv:1003.1852 [hep-ph]].

\bibitem{Albacete:2014fwa} 
  J.~L.~Albacete and C.~Marquet,
  Prog.\ Part.\ Nucl.\ Phys.\  {\bf 76} (2014) 1.
  [arXiv:1401.4866 [hep-ph]].
  
  \bibitem{Adare:2011sc}
  A.~Adare {\it et al.}  [PHENIX Collaboration],
  Phys.\ Rev.\ Lett.\  {\bf 107} (2011) 172301.
  [arXiv:1105.5112 [nucl-ex]].

\bibitem{Braidot:2010zh}
  E.~Braidot [STAR Collaboration],
  arXiv:1005.2378 [hep-ph].

\bibitem{Marquet:2007vb}
  C.~Marquet,
  Nucl.\ Phys.\ A {\bf 796} (2007) 41.
 [arXiv:0708.0231 [hep-ph]].

\bibitem{Boer:2011fh}
  D.~Boer {\it et al.},
  arXiv:1108.1713 [nucl-th].

\bibitem{Albacete:2012rx}
  J.~L.~Albacete, J.~G.~Milhano, P.~Quiroga-Arias and J.~Rojo,
  Eur.\ Phys.\ J.\ C {\bf 72} (2012) 2131.
  [arXiv:1203.1043 [hep-ph]].

\bibitem{Eskola:2002yc}
  K.~J.~Eskola, H.~Honkanen, V.~J.~Kolhinen, J.~w.~Qiu and C.~A.~Salgado,
  Nucl.\ Phys.\ B {\bf 660} (2003) 211.
  [hep-ph/0211239].

\bibitem{Cazaroto:2008iy}
  E.~R.~Cazaroto, F.~Carvalho, V.~P.~Goncalves and F.~S.~Navarra,
  Phys.\ Lett.\ B {\bf 671} (2009) 233.
  [arXiv:0805.1255 [hep-ph]].
    
\bibitem{Accardi:2011qh}
  A.~Accardi, V.~Guzey and J.~Rojo,
  arXiv:1106.3839 [hep-ph].

\bibitem{Balitsky:1995ub}
  I.~Balitsky,
  Nucl.\ Phys.\ B {\bf 463} (1996) 99.
  [hep-ph/9509348].

\bibitem{Kovchegov:1999yj}
  Y.~V.~Kovchegov,
  Phys.\ Rev.\ D {\bf 60} (1999) 034008.
  [hep-ph/9901281].
  
  \bibitem{Gardi:2006rp}
  E.~Gardi, J.~Kuokkanen, K.~Rummukainen and H.~Weigert,
  Nucl.\ Phys.\ A {\bf 784} (2007) 282.
  [hep-ph/0609087].
  
  \bibitem{Kovchegov:2006vj}
  Y.~V.~Kovchegov and H.~Weigert,
  Nucl.\ Phys.\ A {\bf 784} (2007) 188.
  [hep-ph/0609090].
  
  \bibitem{Balitsky:2006wa}
  I.~Balitsky,
  Phys.\ Rev.\ D {\bf 75} (2007) 014001.
  [hep-ph/0609105].
  
 \bibitem{Balitsky:2008zza}
  I.~Balitsky and G.~A.~Chirilli,
  Phys.\ Rev.\ D {\bf 77} (2008) 014019.
  [arXiv:0710.4330 [hep-ph]].

\bibitem{Eskola:2009uj}
  K.~J.~Eskola, H.~Paukkunen and C.~A.~Salgado,
  JHEP {\bf 0904} (2009) 065.
  [arXiv:0902.4154 [hep-ph]].

\bibitem{deFlorian:2011fp}
  D.~de Florian, R.~Sassot, P.~Zurita and M.~Stratmann,
  Phys.\ Rev.\ D {\bf 85} (2012) 074028.
  [arXiv:1112.6324 [hep-ph]].
  
 \bibitem{Giele:1998gw}
  W.~T.~Giele and S.~Keller,
  Phys.\ Rev.\ D {\bf 58} (1998) 094023.
  [hep-ph/9803393].

\bibitem{Ball:2010gb}
  R.~D.~Ball {\it et al.}  [NNPDF Collaboration],
  Nucl.\ Phys.\ B {\bf 849} (2011) 112
   [Erratum-ibid.\ B {\bf 854} (2012) 926]
   [Erratum-ibid.\ B {\bf 855} (2012) 927].
  [arXiv:1012.0836 [hep-ph]].

\bibitem{Ball:2011gg}
  R.~D.~Ball, V.~Bertone, F.~Cerutti, L.~Del Debbio, S.~Forte, A.~Guffanti, N.~P.~Hartland and J.~I.~Latorre {\it et al.},
  Nucl.\ Phys.\ B {\bf 855} (2012) 608.
  [arXiv:1108.1758 [hep-ph]].
 
\bibitem{Watt:2012tq}
  G.~Watt, R.~S.~Thorne,
  JHEP {\bf 1208} (2012) 052.
  [arXiv:1205.4024 [hep-ph]].
  
\bibitem{Watt:2013oha}
  B.~J.~A.~Watt, P.~Motylinski and R.~S.~Thorne,
  arXiv:1311.5703 [hep-ph].

\bibitem{Sato:2013ika}
  N.~Sato, J.~F.~Owens and H.~Prosper,
  arXiv:1310.1089 [hep-ph].
 
\bibitem{Paukkunen:2014zia}
  H.~Paukkunen and P.~Zurita,
  JHEP {\bf 1412} (2014) 100.
  [arXiv:1402.6623 [hep-ph]].
 


\bibitem{Armesto:2002ny}
  N.~Armesto,
  Eur.\ Phys.\ J.\ C {\bf 26} (2002) 35.
  [hep-ph/0206017].

\bibitem{Kowalski:2003hm}
  H.~Kowalski and D.~Teaney,
  Phys.\ Rev.\ D {\bf 68} (2003) 114005.
  [hep-ph/0304189].

\bibitem{DeJager:1987qc}
  H.~De Vries, C.~W.~De Jager and C.~De Vries,
  Atom.\ Data Nucl.\ Data Tabl.\  {\bf 36} (1987) 495.

\bibitem{Albacete:2009fh}
  J.~L.~Albacete, N.~Armesto, J.~G.~Milhano and C.~A.~Salgado,
  Phys.\ Rev.\ D {\bf 80} (2009) 034031.
  [arXiv:0902.1112 [hep-ph]].
  
  

\bibitem{McLerran:1997fk}
  L.~D.~McLerran and R.~Venugopalan,
  Phys.\ Lett.\ B {\bf 424} (1998) 15.
  [nucl-th/9705055].

\bibitem{Albacete:2010sy}
  J.~L.~Albacete, N.~Armesto, J.~G.~Milhano, P.~Quiroga-Arias and C.~A.~Salgado,
  Eur.\ Phys.\ J.\ C {\bf 71} (2011) 1705.
  [arXiv:1012.4408 [hep-ph]].


\bibitem{Albacete:2015xza}
  J.~L.~Albacete,
  Nucl.\ Phys.\ A {\bf 957} (2017) 71.
  [arXiv:1507.07120 [hep-ph]].

\bibitem{Matt}
We are especially grateful to Matt Lamont for providing the projected error bars to us.


\bibitem{Martin:2009iq}
  A.~D.~Martin, W.~J.~Stirling, R.~S.~Thorne and G.~Watt,
  Eur.\ Phys.\ J.\ C {\bf 63} (2009) 189.
  [arXiv:0901.0002 [hep-ph]]. 

\bibitem{Armesto:2015lrg} 
  N.~Armesto, H.~Paukkunen, J.~M.~Pen'n, C.~A.~Salgado and P.~Zurita,
  Eur.\ Phys.\ J.\ C {\bf 76}, no. 4, 218 (2016).
  [arXiv:1512.01528 [hep-ph]].
  
\bibitem{Kugeratski:2005ck}
  M.~S.~Kugeratski, V.~P.~Goncalves and F.~S.~Navarra,
  Eur.\ Phys.\ J.\ C {\bf 46} (2006) 413.
  [hep-ph/0511224].
  
  \bibitem{Kowalski:2007rw}
  H.~Kowalski, T.~Lappi and R.~Venugopalan,
  Phys.\ Rev.\ Lett.\  {\bf 100} (2008) 022303.
  [arXiv:0705.3047 [hep-ph]].
  
  \bibitem{Kowalski:2008sa}
  H.~Kowalski, T.~Lappi, C.~Marquet and R.~Venugopalan,
  Phys.\ Rev.\ C {\bf 78} (2008) 045201.
  [arXiv:0805.4071 [hep-ph]].
  
\end{thebibliography}
\end{document}